\documentclass{aa}
\usepackage{natbib,graphicx,hyperref} 
\usepackage[varg]{txfonts}
\newcommand{\rsun}{R_{\odot}}

\begin{document}

\title{The relationship between flux emergence and subsurface toroidal magnetic flux}
\author{ R.H. Cameron\inst{1} and J. Jiang\inst{2}} 
\institute{$^1$Max-Planck-Institut f{\"u}r Sonnensystemforschung\\
 $^2$ School of Space and Environment, Beihang University, Beijing 100083, P. R. China}
\date{Received ; accepted}

\titlerunning{Flux emergence and subsurface toroidal flux}
\authorrunning{Cameron \& Jiang}

\abstract
{}
{The 1-D mean-field equation describing the evolution of the subsurface toroidal field can be used together
with the observed surface radial field to model the subsurface toroidal flux density.
We  aim to test this model and
  determine the relationship between the observationally inferred surface toroidal field (as a proxy for flux emergence),
  and the modelled subsurface toroidal flux density.}
{We use a combination of sunspot area observations, the surface toroidal field inferred from WSO line-of-sight magnetic field
  observations, and compare with the results of a 1-D mean-field evolution equation for the subsurface toroidal field,
  driven by the observed radial field from the National Solar Observatory/Kitt Peak and SOLIS
 observations.}
{We derive calibration curves relating the subsurface toroidal flux density to the observed surface toroidal field strengths
  and sunspot areas. The calibration curves are for two regimes, one corresponding to ephemeral region emergence
  outside of the butterfly wings,
  the other to active region emergence in the butterfly wings.
  We discuss this in terms of the size and vertical velocity associated with the two types of flux emergence.}
{}


\maketitle

\section{Introduction}
The most visible manifestation of the solar dynamo are sunspots. These appear at latitudes of about $\pm 30^{\circ}$ at
the beginning of a solar activity cycle, and progressively closer to the equator as the solar cycle progresses to latitudes
of about $\pm 8^{\circ}$ \citep{Spoerer89} 11 years later \citep{Schwabe1849}. Sunspots appear as part of the flux emergence
process, where magnetic field rises through the photosphere from below in bipolar active regions
\citep[for a review see][]{2014LRSP...11....3C} .
The east-west orientation of the two polarities of  active regions mostly (about 95\% for large active regions)
follows Hale's law \citep{1919ApJ....49..153H}, indicating that they originate from the subsurface toroidal field of
the global solar dynamo \citep[for a review of dynamo theory see][]{2014ARA&A..52..251C, 2010LRSP....7....3C}.

The solar dynamo is thought to be an alpha-omega dynamo. This means the subsurface toroidal
magnetic field is mainly generated by differential rotation acting on poloidal magnetic field. In addition to the generation by
differential rotation, the toroidal field will be acted on by the large-scale meridional flow and small-scale turbulent convective
eddies. The effect of these flows in combination on the azimuthal toroidal flux can be described using mean-field
magnetohydrodynamics \citep[for example, see][]{Krause80}. The well-known 1-D mean-field equation for the toroidal field  will be discussed in Section~2.
The ability of this 1-D equation to describe the evolution of the toroidal field
can be tested in conjunction with a few additional assumptions. These assumptions are that the poloidal field corresponds to
the radial field at the surface, and that rate at which systematically east-west oriented flux emerges is related to the amount of
subsurface toroidal field. These latter assumptions are part of the Babcock-Leighton model (some variants of which fall within the
mean-field formalism). 

In this paper we investigate whether the 1-D mean field equation with the assumption that the observed surface radial field determines the poloidal
flux provides a good description of the field evolution by comparing the observed toroidal flux in the photosphere
as a measure  of flux emergence \cite{2018A&A...609A..56C},
as inferred from Wilcox Solar Observatory full-disk observations, with the subsurface toroidal field predicted from the evolution equation for
subsurface toroidal field and surface radial field measurements inferred from the Kitt Peak National Solar Observatory (NSO) and SOLIS observations.  
The comparison is done both for active region emergences in the butterfly wings and ephemeral region emergence in the quiet Sun.

Flux emergence occurs in discrete events, for example the emergence of a large active region will occur over a few days and a
range in latitude of a few degrees and up to 10 degrees in longitude. There is, from the point of view of a mean-field model,
a considerable randomness in the details of the latitudes and timing of the flux emergence events.  
This will affect our comparison -  we expect some scatter due to the random nature of the flux emergence process. However,
we will show that taking yearly averages and latitudinal averages over several degrees is sufficient for the systematic component
to be well observed.

The paper is organized as follows, in Section 2 we  present the WSO and NSO/SOLIS observations and the model.  We compare
the results of the model driven with the surface observations of the radial surface field with the observed toroidal field in
Section 3. We present our conclusions in Section 4.

\section{Methods}
The surface toroidal (here meaning the azimuthally-averaged azimuthal) magnetic field corresponds to flux
emergence and can be inferred from observations \cite{2018A&A...609A..56C}.
As explained in that paper the observed surface toroidal field results from flux emergence, where horizontal
magnetic field is carried across the solar surface. The amount of toroidal field observed at the surface is proportional
to the flux emergence rate, with the constant of proportionality depending on the geometric properties of the
emergence events and the rise velocity of the flux through the photosphere. The geometric properties and rise
velocity differ substantially between active regions and ephemeral regions, and hence we will analyse the
activity wings (where active regions emerge) and outside the activity wings (where only ephemeral regions emerge) separately.
    
\begin{figure}
  \resizebox{0.95\hsize}{!}{\includegraphics{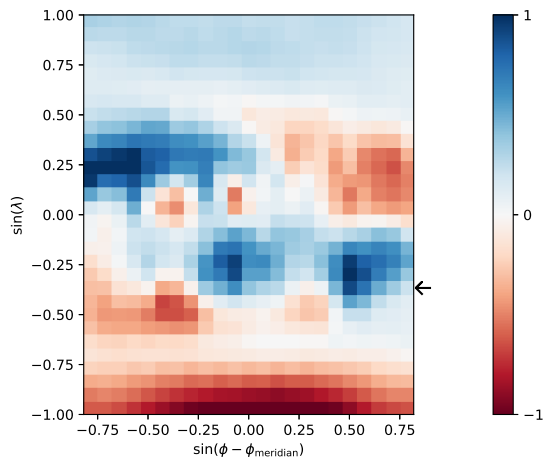}}
  \caption{Yearly averaged (September 6 2015 to September 5 2016) line-of-sight magnetic field as a
    function of latitude and longitude relative to the central meridian $(\phi-\phi_{\mathrm{meridian}})$
    The arrow indicates the latitude used in Figure~\ref{fig:fit} where we illustrate the fitting procedure.}
  \label{fig:deep}
\end{figure}

\begin{figure}
  \resizebox{0.95\hsize}{!}{\includegraphics{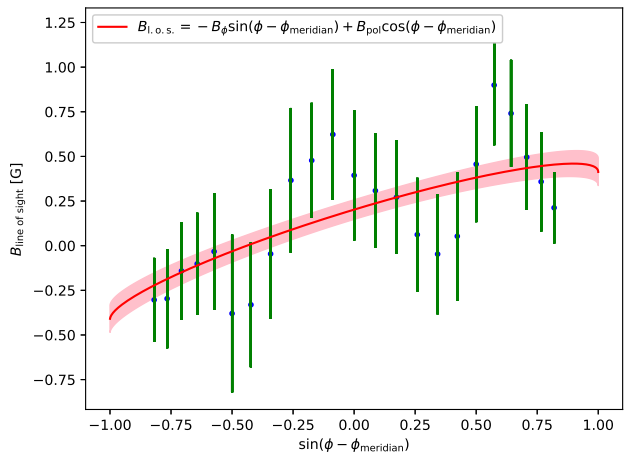}}
  \caption{Yearly averaged (September 6 2015 to September 5 2016) line-of-sight magnetic field at latitude $-21^{\circ}$
    (blue dots). The green-curves show the error estimate based on the scatter in the daily value going into the 12-month
    averaging. The red curve shows the fit
    $B_{\mathrm{l.o.s.}}=-B_{\phi} \sin(\phi-\phi_{\mathrm{meridian}})+B_{\mathrm{pol}} \cos(\phi-\phi_{\mathrm{meridian}})$, and the pink shaded
    region shows the 1-$\sigma$ error estimates for the parameter fits.}
  \label{fig:fit}
\end{figure}

\begin{figure}
  \resizebox{0.95\hsize}{!}{\includegraphics{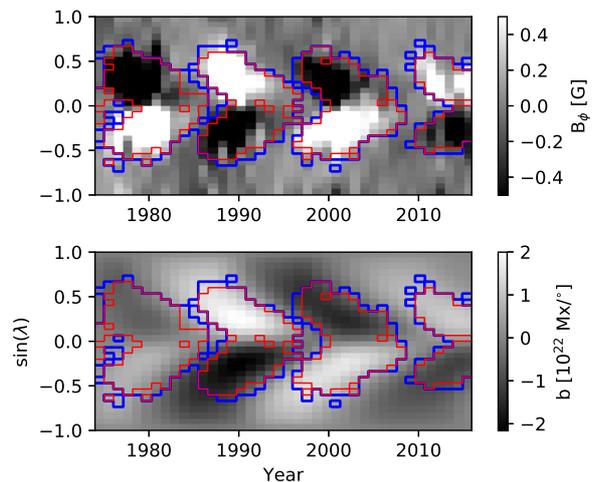}}
  \caption{Upper panel: Azimuthally averaged surface toroidal magnetic field ($B_{\phi}$) as a function time and latitude,
    saturated at $\pm 0.5$~G.
    The maps are inferred from WSO full-disk magnetograms as described in the text. The blue curve outlines the pixels
    (times and latitudes) where any sunspots were reported in the RGO/SOLIS sunspot records, the red curves outline the
    region where the  average coverage by sunspots exceeds 0.1\%  micro Hemispheres over the pixel.
    Lower panel: subsurface toroidal flux density based on the model and observed surface radial field. The red and blue curves
    are the same as in the upper panel. We use the interior of the red curve as the butterfly wings and the exterior of the
    blue contour as the quiet-Sun. }
  \label{fig:bf_qs_ar}
\end{figure}

The procedure for determining the surface toroidal field is set out
in \cite{1978PhDT.........4D}, \cite{1979SoPh...61..233D} and \cite{2018A&A...609A..56C}. The starting point is
the 40 years of WSO full disk magnetograms. In the current study we divide the magnetograms into 12-month series
from the 6th of September each year to the 5th of September of the following year 
(the dates chosen to correspond to the maximum of the solar B angle).  
The magnetograms for each 12 months are used to create the average magnetogram as
seen from the Earth, and an example the average magnetogram is shown in Figure~\ref{fig:deep}. For each
year and latitude, we then have a series of measurements of the line-of-sight field as a function of longitude with
respect to the central meridian, an example of which is shown in Figure~\ref{fig:fit}. We fit the 12-month averaged observed
line-of-sight field for each latitude using
\begin{eqnarray}
  B_{\mbox{los}}=-B_\phi \sin(\phi-\phi_{\mbox{meridian}}) + B_{\mathrm{pol}} \cos(\phi-\phi_{\mbox{meridian}})
\end{eqnarray}
where $B_\phi$ is the inferred toroidal field and $B_{\mathrm{pol}}$ the inferred line-of-sight component of the poloidal field.
The error bars of each fit are inferred from the scatter in the data as a function of $\phi-\phi_{\mbox{meridian}}$.
The upper panel of Figure~\ref{fig:bf_qs_ar} shows the butterfly map of the toroidal field constructed in this way.

In addition to the observed surface toroidal field we need to model the subsurface toroidal field.
For this we use the 1-D mean field equation describing the evolution of the toroidal field under the assumption that
winding up of the field differential rotation (the omega effect) dominates the alpha effect in the production of toroidal flux.
This is likely to be the case in the Sun \cite[for example see][]{2010LRSP....7....3C}. We use the form of this equation
given in \cite{2017A&A...599A..52C} (which is appropriate when the effect of the poloidal field is determined by the surface radial field):

\begin{eqnarray}
  \frac{\partial b}{\partial t}
  &=& \sin\theta\,
  R_\odot^2 B_{r}\mid_{R_{\odot}} \epsilon \left( \Omega_{R_\odot} -
  \Omega_{R_{\rm NSSL}} \right) \nonumber \\
  & &  -\frac{\partial \Omega_{R_{\rm NSSL}}}{\partial \theta}
  \int_0^{\theta} \sin\theta\, R_{\odot}^2 B_{r}\mid_{R_{\odot}}
  \mathrm{d}\theta   \nonumber \\ 
  & & -\frac{1}{\rsun} \frac{\partial}{\partial\theta} 
  \left( V_\theta  \, b \right)
  + \frac{\eta_0}{\rsun^2} \frac{\partial}{\partial\theta}\left[
    \frac{1}{\sin\theta}\frac{\partial}{\partial\theta}
    (\sin\theta\,b)\right] \,.
\label{eq:btor_1}
\end{eqnarray}
where $b=\int B_{\phi} r \mathrm{d}r$ is the subsurface toroidal flux density, $\Omega_{R_\odot}$ and $\Omega_{R_{\rm NSSL}}$
are the solar rotation rates at the solar surface and at the base of the near-surface shear layer,
respectively, $\epsilon$ is a number between 0 and 1 ($\epsilon=1$ if the field lines turn over at the
base of the near-surface shear layer, and is smaller if the field lines turn over deeper where the solar rotation
rate is smaller),
$V_{\theta}$ is the radially averaged $\theta$ component of subsurface meridional flow (where the
average is weighted by the radial distribution of the toroidal flux), and can in principle include both the
bulk flow and the effect of latitudinal  magnetic pumping,
$B_r\mid_{R\odot}$ is the observed radial field at the solar
surface, and $\eta_0$ is the turbulent magnetic diffusivity. For the subsurface meridional flow,
we assume the form $V_{\theta}=V_0 \sin(2\theta)$ where $V_0$ is a free parameter of the model.
The meaning of the parameters and the sensitivity of the calculations to those parameters is discussed
in \cite{2017A&A...599A..52C} where a parameter study was made.
In this paper we concentrate on the case with $V_0=2.5$~m/s and $\eta_0=100$~km$^2$/s
and $\epsilon=1$ (we found similar results with a range of other values of the parameters).

We comment that Equation~\ref{eq:btor_1} corresponds to one of the two equations of the updated
Babcock-Leighton model, exactly because the evolution of the toroidal field in the Babcock-Leighton
model is that of the mean-field description of an alpha-omega dynamo. The equation is also that of
a flux transport dynamo when the latitudinal velocity $V_\theta$ plays a substantial role. This 
flow corresponds to a combination of the equatorward subsurface return flow
(the counterpart to the well-observed surface meridional flow) and the subsurface latitudinal pumping
which is not observationally constrained. For a 1-D model these are weighted radial averages with a
weighting corresponding to the unknown radial distribution of the subsurface toroidal field.
These considerations result in the flow being currently poorly constrained by observations. 
The flow is expected, based on symmetry arguments, to vanish at the equators and poles (on average) which
is the basis for the assumption of the $\sin(2\theta)$ latitudinal form. The amplitude is a free parameter
of the model and the value chosen has previously been used to qualitatively reproduce the butterfly diagram.
The sign of $V_{\theta}$ corresponds to equatorial propagation. 

\begin{figure}
  \resizebox{0.95\hsize}{!}{\includegraphics{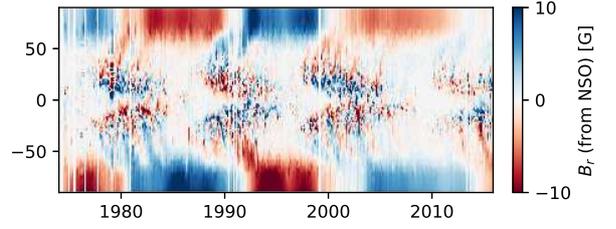}}
  \caption{Azimuthally averaged radial field from NSO/Kitt Peak and SOLIS observations
    as a function of latitude and time.}
  \label{fig:radial}
\end{figure}

We drive the evolution of the subsurface toroidal field using observations: 
Equation~\ref{eq:btor_1} requires knowledge of the surface radial field for which we use
NSO/Kitt Peak and SOLIS observations as presented by \cite{2017SSRv..210...77P}, shown in
Figure~\ref{fig:radial}.  The resulting  model subsurface toroidal field 
mapped to the spatial and temporal resolution of the observed surface toroidal field
is shown in the lower panel of Figure~\ref{fig:bf_qs_ar}.

\section{Results}

\begin{figure}
  \resizebox{0.95\hsize}{!}{\includegraphics{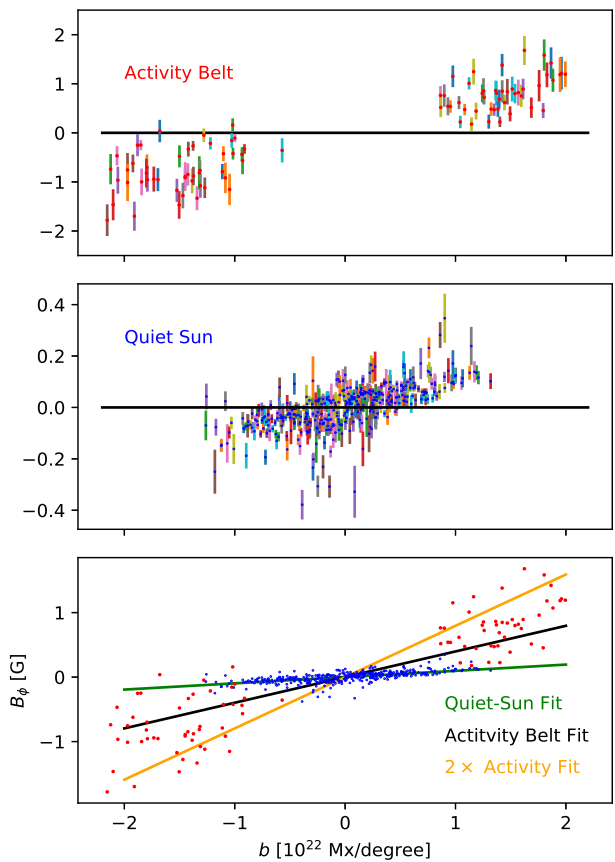}}
  \caption{Each point in the three panels shows the pixel-by-pixel comparison of the subsurface toroidal flux density
    from the model, $b$, and the observed surface toroidal field, $B_{\phi}$, for cycles 22, 23 and 24.
    We excluded cycle 21 from this analysis as it is heavily affected by the initial condition $b=0$
    The upper panel shows the results for the butterfly wings where flux emergence includes active region emergences, the
    middle panel shows results for the quiet-Sun where the emergences are in the form of ephemeral regions, and the bottom panel
    shows both types of regions. The first two panels also show the error bars on the observed surface toroidal field obtained
    from the fitting procedure, in different colours to make them easier to distinguish. In the bottom panel we also
    show fits for both the quiet-Sun and butterfly wing data.}
  \label{fig:ans}
\end{figure}

The observed surface toroidal field and subsurface toroidal flux density are shown in Figure~\ref{fig:bf_qs_ar}.
We have divided the two images into the butterfly wings and quiet-Sun regions based on sunspot
coverage\footnote{The sunspot area data was obtained from https://solarscience.msfc.nasa.gov/greenwch.shtml}
We now consider a pixel by pixel comparison of the toroidal surface field and subsurface flux density for both the
butterfly wings and the quiet-Sun, with the result shown in Figure~\ref{fig:ans}.

Qualitatively, the upper two panels of Figure~\ref{fig:ans} show that in both  butterfly wings and in outside the butterfly wings
the variability of the surface toroidal field is partly captured by the model - a clear approximately linear trend is visible.
We also see that the butterfly wings mostly correspond to subsurface toroidal flux
densities of over $10^{22}$~Mx/degree. As a threshold subsurface flux density for the formation of sunspots, this should be compared
to the range of fluxes of bipolar active regions with sunspots of $5\times10^{21}$ to $3\times10^{22}$
\citep[Table 5.1 of][]{2008ssma.book.....S} and to a spatial scale of a few degrees. The threshold of $10^{22}$~Mx/degree is thus
higher than what is strictly necessary to account for the amount of flux which emerges, but only by a factor of a few, and might reflect
the fact that not all the flux available beneath the surface emerges during each emergence event. The next point to note is that
the average observed surface toroidal field increases with the subsurface toroidal flux density, but there is a large amount of scatter.
Quantitatively we chose to fit the relationship for the butterfly wings with a function of the form $B_{\phi}=a_{\mathrm{wings}} b$ where
$a_{{\mathrm{wings}}}=4.0 \times 10^{-23} \pm 0.1 \times 10^{-23}$~G/(Mx/degree) is the fitted parameter.
The quality of the fit is poor, $\chi^2=501$ for the $93$ data pairs (with $\chi^2/(93-1)=5.4$).
This indicates that the model does not fully explain the surface toroidal field. This presumably reflects a combination of
problems with the model and the fact that the apparently partly stochastic flux emergence introduces a substantial randomness
into $B_{\phi}$.

For the points in the butterfly wings we also performed a fit to 
\begin{eqnarray}
  B_{\phi} \mathrm{sign}(b)= a \mid b \mid +c
  \label{eqn:fit2}
\end{eqnarray}  
where $a$ and $c$ are
the two fit parameters. The idea was to test if the linear fit mainly reflects that the model gets the sign
of $b$ correct (i.e. the same sign as $B_{\phi}$). Fitting to Equation~\ref{eqn:fit2} allows a non-zero $c$ even in the
case where the model gets the sign correct. However we found that the fit is
$a=3.9\times 10^{-23} \pm 0.5 \times 10^{-23}$~G/(Mx/degree)
and $c=0.04\times 10^{-23} \pm 0.6 \times 10^{-23}$~G/(Mx/degree), so $c$ is small and consistent with 0. Hence, we
conclude that the linear fit does not simply reflect the model getting the sign of the $b$ correct.

In the case of the quiet-Sun,  the underlying subsurface toroidal flux density is
below $1.25 \times 10^{22}$~Mx/degree, and the fit of the surface toroidal field is $B_{\phi}=a_{\mathrm{qs}} b$
with $a_{{\mathrm{qs}}}=0.97 \times 10^{-23} \pm 0.015 \times 10^{-23}$~G/(Mx/degree).
Hence the range of subsurface toroidal fluxes
in the quiet-Sun partly overlaps with that of the butterfly wings. The quality of the fit is again poor, with
$\chi^2=2964$ for $382$ data pairs (with $\chi^2/(382-1)=7.7$), again presumably reflecting a combination of problems with the model and
the random component to flux emergence.

The fits for the quiet-Sun and butterfly wings are very different,
$a_{{\mathrm{qs}}}=0.97 \times 10^{-23} \pm 0.015 \times 10^{-23}$~G/(Mx/degree) 
versus $a_{{\mathrm{wings}}}=4.0 \times 10^{-23} \pm 0.1 \times 10^{-23}$~G/(Mx/degree). A substantial part of this is
presumably related to the rise speed of the ephemeral regions and active regions through the photosphere. As discussed
in \cite{2018A&A...609A..56C}, the surface toroidal field corresponds to horizontal field being carried through the
surface during flux emergence. The impact of the horizontal field depends on how long it takes to traverse the
photosphere, which is inversely proportional to the vertical velocity at the surface. Ephemeral regions rise through the
photosphere more quickly, and thus we expect $a_{{\mathrm{qs}}}$ to be smaller than $a_{{\mathrm{wings}}}$. The degree to which
we expect $a_{{\mathrm{qs}}}$ to be smaller is difficult to estimate because not all the flux which emerges in the butterfly wings
will be in the form of active regions. The point of this paragraph is that the difference between
$a_{{\mathrm{qs}}}$ and $a_{{\mathrm{wings}}}$ does not necessarily indicate a difference in the subsurface dynamics of the two classes
of emergences.

The surface toroidal field corresponding to a given subsurface flux density can then be modelled as
\begin{eqnarray}
  B_{\phi}=\left\{ \begin{array}{ll}
    a_{\mathrm{qs}} \times b  & \mathrm{if } \mid b \mid <10^{22}~\mathrm{mx/degree}\\
    a_{\mathrm{wings}} \times b &   \mathrm{otherwise}
  \end{array}
\right.
\end{eqnarray}

In order to relate the subsurface toroidal flux to the sunspot number, we apply a threshold found above of $10^{22}$~Mx/degree.
We then integrate the flux above this threshold in each hemisphere as a function of time and compare this with the
sunspot area coverage in each hemisphere. The model has no sunspots
appearing during cycle 21 because we began from the (obviously incorrect) initial condition that $b=0$ in 1974 -- the model
has a memory of the initial condition of at least one cycle. The maxima of cycles 22, 23 and 24 are partly reproduced.

\begin{figure}
  \resizebox{0.95\hsize}{!}{\includegraphics{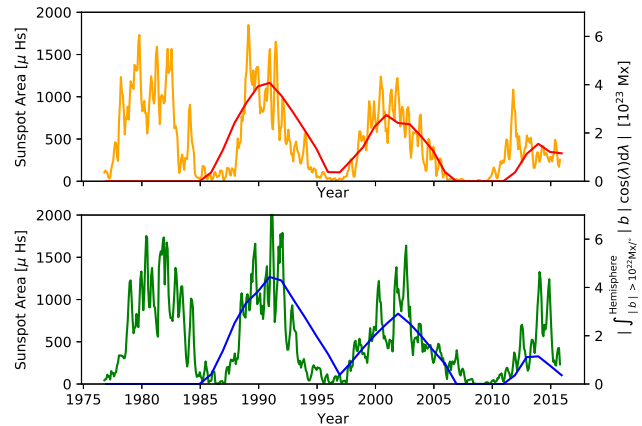}}

  \caption{The sunspot coverage in the northern (southern) hemisphere as a function
    of time in orange (green). Also shown is the integral of the subsurface toroidal flux density
    in each hemisphere where it exceeds $10^{22}$~Mx/degree.}
  \label{fig:SSN}
\end{figure}

\section{Conclusion}

In this paper we investigated the relationship between the surface toroidal field (and sunspot area coverage) as
a function of the subsurface toroidal field as predicted by the surface radial field measurements applied to the
1-D mean-field evolution equation for the subsurface toroidal flux density with a poloidal field derived
from the surface radial magnetic field observations.

We found that the fit between the model subsurface flux density and the surface toroidal field differs in the
quiet-Sun and in the butterfly wings, which we suggest might be because of the different speeds at which ephemeral regions and
active regions emerge through the photosphere. In both cases we found that the quality of the fit as measured by the
$\chi^2$ statistic was relatively poor, possibly because flux emergence introduces a random component.

Our results allow the subsurface toroidal flux density of the model to be quantitatively
related to observable quantities such as sunspot areas and surface toroidal fields.

\begin{acknowledgements}
JJ acknowledges the support by the National Science Foundation of
China (grant Nos. 11873023, 11522325) and by the Fundamental
Research Funds for the Central Universities of China.
  
Wilcox Solar Observatory data used in this study was obtained via the web site http://wso.stanford.edu, courtesy of J.T. Hoeksema.

NSO/Kitt Peak data used here are produced cooperatively
by NSF/NOAO, NASA/GSFC, and NOAA/SEL. This work utilizes SOLIS data
obtained by the NSO Integrated Synoptic Program (NISP), managed by the
National Solar Observatory, which is operated by the Association of
Universities for Research in Astronomy (AURA), Inc. under a cooperative
agreement with the National Science Foundation.

We thank Manfred Sch{\"u}ssler and two referees for useful comments.
\end{acknowledgements}

\bibliographystyle{aa}
\bibliography{comp_model_sim}

\end{document}